\documentclass[letterpaper,journal]{IEEEtran}
\usepackage{amsmath,amsfonts}
\usepackage{algorithmic}
\usepackage{array}
\usepackage[caption=false,font=normalsize,labelfont=sf,textfont=sf]{subfig}
\usepackage{textcomp}
\usepackage{stfloats}
\usepackage{url}
\usepackage{verbatim}
\usepackage{graphicx}
\def\BibTeX{{\rm B\kern-.05em{\sc i\kern-.025em b}\kern-.08em
    T\kern-.1667em\lower.7ex\hbox{E}\kern-.125emX}}
\usepackage{balance}
\usepackage{glossaries}

\begin{document}

\newacronym{CV-QKD}{CV-QKD}{Continuous-Variable Quantum Key Distribution}
\newacronym{LO}{LO}{local oscillator}
\newacronym{WSS}{WSS}{Wide-Sense Stationarity}
\newacronym{SNU}{SNU}{Shot Noise Units}
\newacronym{SNC}{SNC}{Shot Noise Calibration}
\newacronym{SNR}{SNR}{Signal-to-Noise Ratio}
\newacronym{SA}{SA}{Stationarity-Aware}
\newacronym{TA}{TA}{Time-Agnostic}
\newacronym{SNM}{SNM}{Simplified Noise Model}
\newacronym{WNE}{WDE}{White Dynamics Exploiting}
\newacronym{PSD}{PSD}{Power Spectral Density}
\newacronym{RIN}{RIN}{Relative Intensity Noise}
\newacronym{AR}{AR}{AutoRegressive}
\newacronym{MA}{MA}{Moving Average}
\newacronym{BHD}{BHD}{Balanced Homodyne Detection}
\newacronym{SKR}{SKR}{Secret Key Rate}
\newacronym{TGV}{Time-Gated Variance}{Time-Gated Variance}
\newacronym{QKD}{QKD}{Quantum Key Distribution}

\title{Receiver Noise Calibration in CV-QKD\\ accounting for Noise Dynamics }
\author{Guillaume~Ricard*,
        Yves~Jaouën,
        and~Romain~Alléaume \thanks{Guillaume~Ricard, Yves~Jaouën and Romain~Alléaume  are with Telecom Paris - Institut Polytechnique de Paris, 19 Pl. Marguerite Perey, 91120 Palaiseau, France
        
        Romain Alléaume is also with Inria Saclay, 1 Rue Honoré d'Estienne d'Orves, 91120 Palaiseau, France}
        
\thanks{*Ricard@telecom-paris.fr}}

\markboth{Journal of \LaTeX\ Class Files,~Vol.~18, No.~9, September~2020}%
{How to Use the IEEEtran \LaTeX \ Templates}

\maketitle

\begin{abstract}
 \gls{CV-QKD} relies on accurate noise calibration at the receiver to ensure the security of quantum communication. Traditional calibration methods often oversimplify noise characteristics, neglecting the impact of local oscillator (LO) noise and the critical role of noise spectral properties, which can lead to imprecise \gls{SNC}. Our contributions are threefold: 1) we propose an operational framework for calibration, relying on the notion of stationarity 2) in this framework, we give a method allowing us to derive the optimal calibration duration for a given experiment 3) leveraging our knowledge of noise spectral properties, we introduce a novel \gls{SNC} method. This work also formalizes the calibration procedures, addressing implicit assumptions and providing a better foundation for the certification of \gls{CV-QKD} protocols, of which calibration is a fundamental part. We demonstrate that our improved calibration technique offers higher performance and higher tolerance to receiver imperfections, which can enhance the performance and cost-effectiveness of \gls{CV-QKD} systems.
\end{abstract}

\glsresetall


\section{Introduction}
\IEEEPARstart{S}{ecure} communication via \gls{QKD} guarantees information-theoretic security by exploiting a quantum channel to rigorously bound potential information leakage. Among its various implementations, \gls{CV-QKD} (see, e.g., \cite{laudenbach_continuous-variable_2018} for a comprehensive review), which encodes information in the quadratures of the electromagnetic field, offers notable advantages for integration within existing fiber-optic infrastructures. It relies on similar optical source and detection technologies as classical coherent communication systems, and enables high key generation rates over short to moderate transmission distances.

A central procedure in \gls{CV-QKD} is parameter estimation, which consists of inferring security-related parameters from the measured quadrature data, combined with externally calibrated values. These calibrated parameters correspond to quantities that cannot be directly measured during the actual \gls{CV-QKD} run, such as the contribution of isolated noise sources. Among these, the shot noise level is particularly critical for security, as the protocol relies on monitoring its relative increase (referred to as excess noise : noise in excess of the shot noise) to detect potential eavesdropping. Accurate calibration of the shot noise is therefore essential to guarantee both the security and the operational performance of the system.

\begin{figure}
    \centering
    \includegraphics[width=0.45\textwidth, trim= 0 0 0 0, clip]{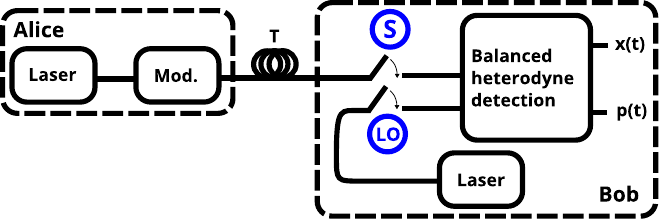}
    
    \caption{Block diagram of our \gls{CV-QKD} experiment. Light generated from Alice’s laser is modulated, propagates through an untrusted channel, and is measured using heterodyne detection with a real local oscillator. Two switches represent calibration settings: Switch \gls{LO} (controls the local oscillator) and Switch S (controls the signal path).}
    \label{fig:block_diag_setup}
\end{figure}

\begin{table}
    \centering
    \renewcommand{\arraystretch}{1.1}
    \setlength{\tabcolsep}{3pt}
    \footnotesize
    \caption{Noise calibration cycle corresponding to Figure~\ref{fig:block_diag_setup}.}
    \begin{tabular}{c|c|c|p{2cm}|p{3cm}}
        
        Calib. step & \gls{LO} & S & Time-domain signal & Variance\\
        \hline
        \hline
        1 & OFF & OFF & $n_{\text{elec}}$ & $\sigma_{\text{elec}}^2(\tau)$ \\ 
        \hline
        2 & ON & OFF & $n_{\text{rec}}$ & $\sigma_{\text{rec}}^2(\tau)$\\ 
        \hline
        3 & ON & ON & 
        $\sqrt{\frac{\eta T}{2}} ( x_A +\epsilon)+ n_{\text{rec}}$
        & $\frac{\eta T}{2} \left( V_{\text{A}}+ \xi_{\text{Alice}}(\tau) \right) + \sigma_{\text{rec}}^2(\tau)$\\
        %\hline
    \end{tabular}

    \label{tab:parameters}
\end{table}

In practice, isolating shot noise from other noise sources at the receiver, such as electronic noise and technical noise introduced by the \gls{LO}, is a challenging task, since only specific operational regimes, summarized in Figure \ref{fig:block_diag_setup} and Table \ref{tab:parameters}, are experimentally accessible. Conventional calibration methods typically assume that shot noise can be determined simply as the difference between the total receiver noise and the measured electronic noise. Such approaches, whether based on “static” calibration techniques~\cite{fossier_field_2009}, frequent recalibrations (once per data block~\cite{pietri_experimental_2024} or even per pulse~\cite{jouguet_experimental_2013}), neglect contributions from \gls{LO} imperfections as well as dynamic effects stemming from colored noise sources. Alternative methods, such as monitoring the \gls{LO} power with a photodiode to infer shot noise via a pre-established linear relationship~\cite{jouguet_field_2012}, suffer from the same limitations. Another strategy, which treats the entire receiver noise as the \gls{SNU}, introduces its own drawbacks: it prevents electronic noise from being treated as a trusted parameter~\cite{zhang_one-time_2020,zhang_continuous-variable_2019}, and still fails to account for dynamic effects related to colored noise. Moreover, recent studies on noise variance dynamics~\cite{brunner_precise_2020,van_der_heide_receiver_2023}, inspired by tools from oscillator frequency stability analysis~\cite{rutman_characterization_1978}, have highlighted the need to explicitly incorporate dynamics into calibration models. Depending on the quality of the receiver and the duration of calibration, neglecting such dynamics can severely affect performance.

To address this, we introduce a novel calibration approach that explicitly characterizes noise variance dynamics. Our contributions are as follows:
\begin{enumerate}
    \item We define an operational framework leveraging stationarity to support calibration;
    \item We propose a method to derive the optimal calibration duration using a novel variance estimator called the \acrshort{TGV}, which allows for estimating the variance of a finite-time process from its \gls{PSD};
    \item We introduce a new \gls{SNC} technique that incorporates noise variance dynamics. It assumes the noise spectral properties are characterized and that receiver excess noise is fully colored, thereby allowing shot noise to be extracted as the white noise component of the receiver noise.
\end{enumerate}
By accounting for noise spectral properties, our method improves the system’s tolerance to imperfections, enabling better performance and the use of cost-effective receivers, thus advancing practical \gls{CV-QKD} implementations.

\section{Noise Calibration and Parameter Estimation in \gls{CV-QKD}}

\subsection{The \gls{WSS} Assumption}

Calibration assumes that noise variances remain ``valid'' over time, which requires a form of stationarity. However, strict stationarity, where all statistical moments are invariant with respect to time shifts, is unrealistic for real-world stochastic processes. Instead, we assume \gls{WSS}, which requires:
\begin{itemize}
    \item A \textbf{constant mean} over time;
    \item A \textbf{well-defined variance};
    \item An \textbf{autocovariance function} that depends only on time differences, not on absolute time.
\end{itemize}

\begin{figure}
    \centering
    \includegraphics[width=0.45\textwidth, trim=0 0 0 0, clip]{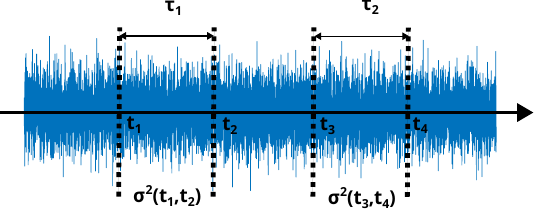}
    \caption{Example of a stochastic process over time. Two windows of duration \(\tau_1\) (from \(t_1\) to \(t_2\)) and \(\tau_2\) (from \(t_3\) to \(t_4\)) are highlighted. The estimated variances are \(\sigma^2(t_1, t_2)\) and \(\sigma^2(t_3, t_4)\).}
    \label{fig:noise}
\end{figure}

A direct consequence of \gls{WSS} is a stable \gls{PSD}, which ensures a consistent frequency-domain representation of noise~\cite{kobayashi_probability_2011}. As a result, the variance, obtained by integrating the \gls{PSD} over frequency, depends solely on the \textbf{observation duration} (denoted by $\tau$), and not on the absolute observation time (denoted by $t$). This is illustrated in Figure~\ref{fig:noise}, estimating variance over a given window (e.g., from \(t_1\) to \(t_2\)) yields a certain value. If we later estimate variance over another window (e.g., from \(t_3\) to \(t_4\)), we need to ensure that the two estimations remain valid and comparable. \gls{WSS} guarantees that variance depends only on the \textbf{duration} of the observation (\(\tau_1\) and \(\tau_2\)) and not on absolute timestamps. Thus if the process is \gls{WSS}, the calibration of noise variance over a duration $\tau$ can be performed by sampling noise calibration data over any other time window of duration $\tau$.

In \gls{CV-QKD}, we leverage this property through characterized dynamics \gls{SNC}, which implicitly requires verifying \gls{WSS} to ensure accurate noise calibration, even when accounting for noise spectral properties.

\subsection{Calibration Cycle}

A standard \gls{CV-QKD} experiment with a locally generated \gls{LO} \cite{qi_generating_2015} follows a cycle composed of three main phases: \textit{calibration}, \textit{measurement}, and \textit{parameter estimation}. The corresponding configurations of the optical switches are illustrated in Figure~\ref{fig:block_diag_setup} and detailed in Table~\ref{tab:parameters}.

\subsubsection{Calibration Phase}

This phase is dedicated to noise characterization, required to calibrate the shot noise variance and define the shot noise unit (\gls{SNU}), which normalizes the parameters used in the security proof, such as excess noise.

At Bob’s receiver, the total detected noise is decomposed into three contributions:
\begin{itemize}
    \item \textbf{Shot noise} ($N_0$): The quantum noise originating from vacuum fluctuations, fundamental to the protocol’s security. Its accurate estimation is crucial.
    \item \textbf{Electronic noise} ($\sigma_{\text{elec}}^2$): Measured by switching off both the signal (S) and the \gls{LO} lasers. It reflects the intrinsic noise of Bob's detection system and is considered trusted.
    \item \textbf{Receiver excess noise} ($\delta_{\text{rec}}$): Often neglected, this term represents technical noise contributions, such as \gls{RIN} from the \gls{LO}, which can otherwise be mistaken for shot noise if not explicitly accounted for. It is therefore included in the following analysis.
\end{itemize}

The calibration phase consists of two sequential steps:
\begin{itemize}
    \item \textbf{Electronic noise calibration}: with both the signal and \gls{LO} switches OFF. The variance of the measured signal $n_{\text{elec}}(t)$ is computed over a duration $\tau$:
    \begin{equation}
    \sigma_{\text{elec}}^2(\tau) = \operatorname{var}(n_{\text{elec}}[t, t+\tau]).
    \end{equation}

    \item \textbf{Full receiver noise calibration}: with the \gls{LO} switch ON and the signal switch OFF. The total noise $n_{\text{rec}}(t)$ includes shot noise and receiver excess noise in addition to electronic noise. Its variance over a duration $\tau$ is:
    \begin{equation}
    \sigma_{\text{rec}}^2(\tau) = \sigma_{\text{elec}}^2(\tau) + N_0 + \delta_{\text{rec}}(\tau).
    \end{equation}
\end{itemize}

Note that in principle, one could use different measurement durations to estimate electronic noise, receiver noise, and signal variance independently. However, such an approach would require very careful control and modeling to ensure consistency and comparability between these estimations during parameter estimation. For practical and conceptual simplicity, it is therefore natural to adopt a single observation duration \(\tau\) across all calibrations. This choice facilitates the interpretation, modeling, and combination of noise variances in the \gls{CV-QKD} protocol.

\subsubsection{Measurement Phase}

In this phase, both the signal and \gls{LO} switches are turned ON. Bob measures the quadrature data, which contains Alice’s modulated signal combined with noise.

\subsubsection{Parameter Estimation Phase}

From the collected measurement (and calibrated) data, key parameters such as channel transmittance and excess noise are estimated. These parameters are then used to assess the channel security and the overall system performance within the framework of the chosen security proof.

\subsection{Signal Measurement and Parameter Estimation}
\label{section:conventional}
Bob measures quadratures $x(t)$ and $p(t)$ using heterodyne detection. The variances provide key information about the received signal :
\begin{equation}
V_B(\tau) = \frac{\eta T}{2}  \left( V_{\text{A}}+ \xi_{\text{Alice}}(\tau) \right) + \sigma_{\text{rec}}^2(\tau),
\end{equation}
\begin{equation}
V_{B|A}(\tau) = \frac{\eta T}{2} \xi_{\text{Alice}}(\tau) + \sigma_{\text{rec}}^2(\tau).
\end{equation}
where:
\begin{itemize}
    \item $\tau$ is the measurement duration.
    \item $V_{B|A}(\tau)$ is the variance of Bob's measured quadrature, conditioned on Alice's sent discrete modulated symbols. This represents the residual noise after accounting for Alice’s modulation, effectively isolating the contributions from channel noise and detection imperfections.
    \item $\eta$ is the detection efficiency (assumed to be known before transmission),
    \item $T$ is the transmittance of the untrusted channel,
    \item $V_{\text{A}}$ is Alice’s modulation variance (assumed to be known before transmission),
    \item $\xi_{\text{Alice}}(\tau)$ is the excess noise rescaled at the channel's input,
    \item $\sigma_{\text{rec}}^2(\tau) = \text{var}(n_\text{rec}[t,t+\tau])$ is the full receiver noise variance,
    
\end{itemize}
The estimated shot noise is obtained as:
\begin{equation}
\hat{N_0}(\tau) = \sigma{_\text{rec}}^2(\tau) - \sigma_{\text{elec}}^2(\tau).
\end{equation}

Measured variances are normalized in \gls{SNU}:
\begin{equation}
V_B^{\text{(SNU)}}(\tau) = \frac{V_B (\tau)}{\hat{N}_0 (\tau)}
\end{equation}

\begin{equation}
V_{B|A}^{\text{(SNU)}}(\tau) = \frac{V_{B|A}(\tau)}{\hat{N}_0 (\tau)}
\end{equation}

From these, we obtain, dropping the time dependencies to simplify notations, estimators for the transmittance $\hat{T}$ and Alice’s excess noise $\hat{\xi}_{\text{Alice}}^{\text{SNU}}$:
\begin{equation}
\label{transmittance}
\hat{T} = \frac{2}{\eta V_A} \left(V_B^{\text{(SNU)}} - V_{B|A}^{\text{(SNU)}}\right)
\end{equation}
\begin{equation}
\label{excess noise}
\hat{\xi}^{(SNU)}_{\text{Alice}} = \frac{2}{{\eta \hat{T}}} \left(V_{B|A}^{\text{(SNU)}} - \frac{\sigma_{\text{rec}}^2}{\hat{N_0}} \right)
\end{equation}

\subsection{Trusted Noise}

\begin{table}

    \centering
    \footnotesize
    \caption{Summary of measured, calibrated, and estimated parameters.}
    \renewcommand{\arraystretch}{1.3}
    \setlength{\tabcolsep}{4pt}
    \begin{tabular}{p{2.9cm}|p{1.2cm}|p{2cm}}
    %\hline
    \textbf{Parameters} & \textbf{Trustable} &  \textbf{Noise Spectral Properties} \\
    \hline 
    \hline
    Electronic noise $\sigma_{\text{elec}}^2$ & Yes &  Partially white  \\
    \hline
    Full receiver noise $\sigma^2_{\text{rec}}$ & Yes &   Partially white \\
    \hline
    Shot noise $N_0$ & Yes &  Fully white \\
    \hline
    Receiver excess noise $\delta_{\text{rec}}$ & Yes &  Not white \\
    \hline
    Excess noise (rescaled at Alice's output) $\xi_{\text{Alice}}$ & No & Partially white \\
    \hline
    Total variance at Bob $V_B$ & N/A &  Partially white  \\
    \hline
    Conditioned variance $V_{B|A}$ & N/A & Partially white \\
    \hline
    Transmittance $T$ & No & N/A \\
    \hline
    Detection efficiency $\eta$ & Yes &  N/A \\

    \end{tabular}

    \label {tab:trusted}
\end{table}

Not all noise components are treated equally in security proofs. Some can be considered trusted in some looser security frameworks \cite{laudenbach_continuous-variable_2018,usenko_feasibility_2010} if they satisfy two conditions:  
\begin{enumerate}    
    \item They are beyond Eve’s control.
    \item They can be precisely calibrated.
\end{enumerate}

For example, electronic noise on Bob’s side is typically considered trusted. It originates from well-characterized components within Bob’s secure area and can be accurately measured (e.g., by blocking the channel output and disabling the local oscillator). The key idea behind trusted noise is that if we are able to precisely calibrate a fraction of the noise for which we know cannot be controlled by Eve, then this noise fraction can be assumed to be independent of  Eve's leakage information, and hence be treated specifically to compute the secure key rate. As a result, even though trusted noise reduces the \gls{SNR}, its impact on the secret key rate is minimal compared to untrusted noise. In contrast, noise that cannot be precisely estimated or that Eve could potentially influence must be considered untrusted, leading to a significant reduction in the achievable key rate.  Table~\ref{tab:trusted} summarizes which noise contributions can reasonably be classified as \emph{trusted}, as well as the expected noise colour.

\glsreset{SNC}
\subsection{Calibration Models}

\begin{table*}
\centering
\caption{Summary of calibration schemes, their stationarity assumptions, and associated shot noise estimators and receiver noise models. }
\begin{tabular}{p{2cm}|p{2cm}|p{5cm}|p{6cm}}
%\hline
\textbf{\gls{SNC} scheme name} & \textbf{Stationarity} & \textbf{Shot noise estimator}& \textbf{Receiver noise model ($\sigma^2_\text{rec})$} \\
\hline
\hline
Uncharacterized dynamics & Unknown & $ \hat{N_0}(\tau  ) = \sigma^2_{rec}(\tau  ) -\sigma^2_{\text{elec}}(\tau)$ & $\sigma^2_\text{rec}(\tau) = N_0(\tau) + \sigma^2_\text{elec}(\tau)$ \\ \hline
Characterized dynamics & WSS & $ \hat{N_0}(\tau_\text{opt}) = \sigma^2_{rec}(\tau_\text{opt}) -\sigma^2_{\text{elec}}(\tau_\text{opt})$ &$\sigma^2_\text{rec}(\tau_\text{opt}) = N_0 + \sigma^2_\text{elec}(\tau_\text{opt}) + \delta_\text{rec}(\tau_\text{opt})$ \\ \hline
Characterized dynamics fully white & WSS & $\hat{N_0} = \sigma^2_\text{rec, white}-\sigma^2_{\text{elec, white}}$& $\sigma^2_\text{rec}(\tau) = N_0 + \sigma^2_{\text{elec, white}} + \sigma^2_\text{rec,not white}(\tau)$\\
%\hline
\end{tabular}
\label{tab:overview}
\end{table*}

Precise calibration of the shot noise level is crucial in \gls{CV-QKD}, as it directly influences both the security and operational performance of the protocol. The most widely employed calibration method relies on a simplified model in which the total receiver noise is assumed to comprise two independent components: shot noise and electronic noise. In this conventional approach, the shot noise variance is estimated by subtracting the electronic noise (LO off) from the total receiver noise (LO on).

However, this standard model implicitly assumes that noise variances remain constant between the calibration and measurement phases, an assumption of stationarity. It further neglects contributions from technical noise sources, such as LO intensity fluctuations, and disregards the impact of colored noise components and calibration duration. Consequently, this oversimplified model can produce inaccurate shot noise estimates, degrading system performance.

To overcome these limitations, we introduce and compare three calibration models :

\begin{itemize}
    \item The \textit{Uncharacterized Dynamics \gls{SNC}} model follows the conventional shot noise estimation procedure. It does not monitor or account for possible temporal variations in noise variance between the calibration and measurement phases, as well as the impact of calibration durations.
    \item The \textit{Characterized Dynamics \gls{SNC}} model retains the same estimation principle but supplements it with a \gls{WSS} hypothesis and characterization of the receiver noise variance dynamics, enabling calibration durations and procedures to be adapted in light of observed noise spectral features. This allows, for example, to choose the optimal calibration duration $\tau_{opt}$ for a given system.
    \item The \textit{Characterized Dynamics, Fully White \gls{SNC}} model incorporates prior knowledge of the noise spectral properties to more accurately estimate the true noise variance, irrespective of the calibration duration.
\end{itemize}

Table~\ref{tab:overview} summarizes the assumptions and features of these calibration models, which are used throughout the remainder of this paper.

\section{Finding the optimal calibration duration}

A natural question arises: within the duration \(\tau_\text{max}\) corresponding to the maximal time during which the signal remains \gls{WSS}, in characterized dynamics \gls{SNC}, is it always better to calibrate for as long as possible? It seems reasonable to think so, as a longer calibration at the same sampling rate provides more samples, improving the statistical precision of variance estimation and thus enhancing performance. However, excessively long calibration also has drawbacks. 

\subsection{\acrshort{TGV} Estimation}

The goal of this section is to quantify the impact of time-limited measurement over noise variance estimate. 

Variance can be expressed for a real-valued process as:
\begin{align}
    \sigma^2 &= \int_{-\infty}^{\infty} S_x(f)df \\
    &= 2  \int_{0}^{\infty}S_x(f)df.
\end{align}
With $S_x(f)$ the signal's \gls{PSD}. 

\begin{figure}
    \centering
    \includegraphics[width=0.27\textwidth, trim= 0 0 0 0, clip]{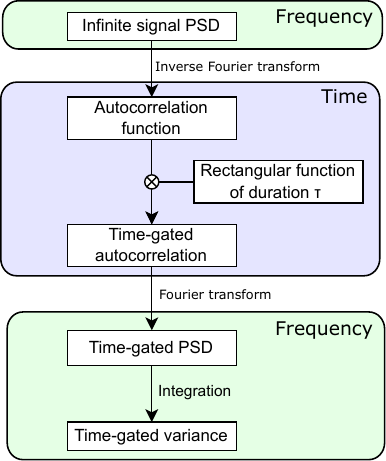}
    \caption{Block diagram of the practical computation of \acrshort{TGV} from the infinite signal \gls{PSD}. For convenience, the signal is inverse Fourier transformed to the time domain, allowing for straightforward windowing. It is then transformed back to the frequency domain for variance estimation through integration.}
    \label{fig:block_diag}
\end{figure}

To account for the finite time of calibration, we modify the variance expression by considering the effect of a finite observation period, through a procedure summarized in figure \ref{fig:block_diag}. The time-gate function, a rectangular function, is defined as:
\begin{equation}
\text{rect}_\tau(t) = 
\begin{cases}
1, & \text{if } |t| \leq \frac{\tau}{2} \\
0, & \text{if } |t| > \frac{\tau}{2}
\end{cases}
\end{equation}

From the Wiener–Khinchin theorem, we can write the Fourier relationship between \gls{PSD} and the autocorrelation function for \gls{WSS} processes \cite{kobayashi_probability_2011}:
\begin{align}
    \mathcal{F}^{-1}(S_x(f))(t) &= \int_{-\infty}^{\infty} S_x(f)e^{2i\pi f t} df \\
    &= R_x(t).
\end{align}

Applying the time gate to the autocorrelation function:
\begin{equation}
    R_{x,\text{gated}}(t,\tau) = R_x(t) \cdot \text{rect}_\tau(t).
\end{equation}

The Fourier transform of this function is:
\begin{align}
    \mathcal{F}(R_{x,\text{gated}}(t,\tau))(f) &= \int_{-\tau/2}^{\tau/2} R_{x,\text{gated}}(t,\tau) e^{-2i\pi f t} dt \\
    &= S_{x,\text{gated}}(f,\tau).
\end{align}

To assess the effect of $\tau$ on the variance estimator, we integrate the new \gls{PSD} over all relevant frequencies to obtain the \acrshort{TGV} estimate.

\begin{align}
\label{sigma gated}
    \sigma^2_{\text{gated}}(\tau) &=  2  \int_{f_\text{min}}^{f_\text{max}}S_{x,\text{gated}}(f,\tau)df,
\end{align}
where the lowest resolvable frequency $f_\text{min} =  1/\tau$ is determined by the total signal duration and the highest meaningful frequency $f_\text{max} = f_s/2$ by the sampling rate $f_s$.

\subsection{Spectral noise Model}
\label{results}

We define our noise model in terms of its \gls{PSD}, distinguishing between electronic noise and full receiver noise contributions (used parameters in table \ref{tab:params_model_noise}). 

\subsubsection{Electronic Noise Model}
The \gls{PSD} of the electronic noise is given by:
\begin{equation}
    S_{\text{elec}}(f) = \sum_{\alpha = -2}^{2} h_{\alpha}^{\text{elec}} f^{\alpha} + A_{\text{peak}}\delta(f - f_{\text{peak}}),
\end{equation}
where:
\begin{itemize}
    \item $h_{\alpha}^{\text{elec}}$ represents typical power-law spectral components of electronic noise.
    \item The Dirac function $A_{\text{peak}}\delta(f - f_{\text{peak}})$ accounts for a parasitic fixed-frequency tone, typically originating from defects in the detection electronics or ADC.
\end{itemize}

\subsubsection{Full receiver noise Model}
\begin{figure}

    \includegraphics[width=0.48\textwidth, trim=0 0 0 0, clip]{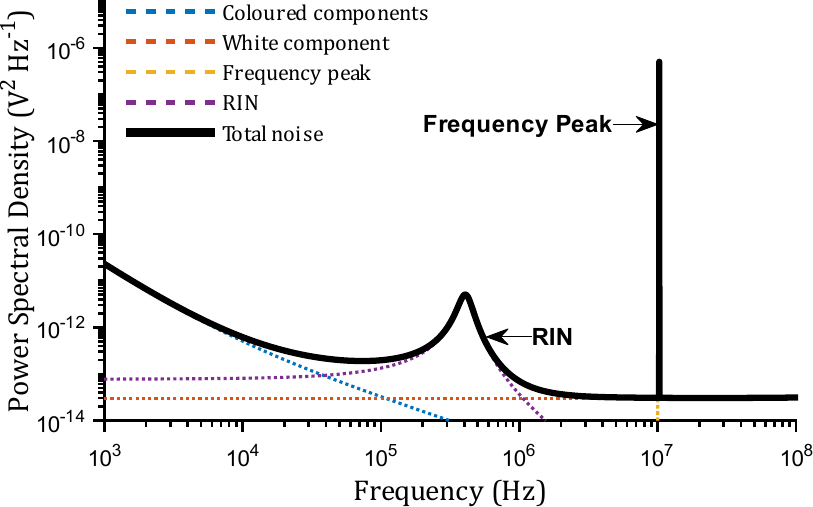}
    \caption{\gls{PSD} of full receiver noise model and its components.}
    \label{fig:PSD_model_loc}
\end{figure}

The full receiver noise \gls{PSD} extends the electronic noise model by incorporating optical noise contributions (displayed in figure \ref{fig:PSD_model_loc}):
\begin{equation}
    S_{\text{rec}}(f) = \sum_{\alpha = -2}^{2} h_{\alpha}^{\text{rec}} f^{\alpha} + A_{\text{peak}} \delta(f - f_{\text{peak}}) + \frac{A_{\text{RIN}}\gamma}{\pi((f - f_{\text{RIN}})^2 + \gamma^2)}
\end{equation}
where:
\begin{itemize}
    \item $h_{\alpha}^{\text{rec}}$ represents the power-law spectral components of full receiver noise, including optical and electronic contributions.
    \item The term $A_{\text{peak}}\delta(f - f_{\text{peak}})$ accounts for the same parasitic tone as in the electronic noise model.
    \item The last term represents the \gls{RIN}, modeled as a Lorentzian function centered at $f_{\text{RIN}}$ with scaling $\gamma$ and height of peak $A_{\text{RIN}}$.
\end{itemize}

\begin{table}
\centering
\caption{Noise model parameters for electronic and full receiver noise contributions.}
\begin{tabular}{p{2cm}|p{2.5cm}|p{2.5cm}}
%\hline
\textbf{Noise Contribution} & \textbf{Electronic Noise Model} & \textbf{Full receiver noise Model} \\ \hline
\hline
$h_{-2}$ & \( 2 \times 10^{-5} \) & \( 2.005 \times 10^{-5} \) \\ %\hline
$h_{-1}$ & \( 3 \times 10^{-10} \) & \( 3.0 \times 10^{-9} \) \\ %\hline
$h_0$ & \( 7 \times 10^{-16} \) & \( 3 \times 10^{-14} \) \\ %\hline
$h_1$ & \( 6 \times 10^{-24} \) & \( 6 \times 10^{-24} \) \\ %\hline
$h_2$ & \( 4 \times 10^{-32} \) & \( 4 \times 10^{-32} \) \\ %\hline
$f_{\text{RIN}}$ & N/A & \( 4 \times 10^5 \) \\ %\hline
$A_\text{RIN}$ & N/A & \( 5 \times 10^{-12} \) \\ %\hline
$\gamma$ & N/A & \( 1 \times 10^5 \) \\ %\hline
$f_{\text{peak}}$ & \( 1 \times 10^7 \) & \( 1 \times 10^7 \) \\ %\hline
$A_\text{peak}$ & \( 5 \times 10^{-7} \) & \( 5 \times 10^{-7} \) \\ %\hline
\end{tabular}

\label{tab:params_model_noise}
\end{table}

\subsection{Time-Gated calibrations}

Let’s define the calibration sequence as follows:

First, the electronic noise is calibrated for a duration $\tau$, yielding the \acrshort{TGV} $\sigma^2_{\text{elec}}(\tau)$. 

Then, the full receiver noise is measured with the \gls{LO} turned on, for a duration $\tau$:

\begin{equation} \sigma^2_{\text{rec}}(\tau) = N_0 + \sigma^2_{\text{elec}}(\tau) + \delta_{\text{rec}}(\tau) \end{equation}

Here, $N_0$ is the true shot noise variance that we are trying to estimate (does not depend on time since it's white noise), $\sigma^2_{\text{elec}}(\tau)$ is the electronic noise variance if it were calibrated over duration $\tau$, and $\delta_{\text{rec}}(\tau)$ represents additional noise contributions present in the receiver noise but not in the electronic noise, that are usually wrongly attributed to shot noise variance.

\begin{figure}
    \includegraphics[width=0.48\textwidth, trim= 0 0 0 0, clip]{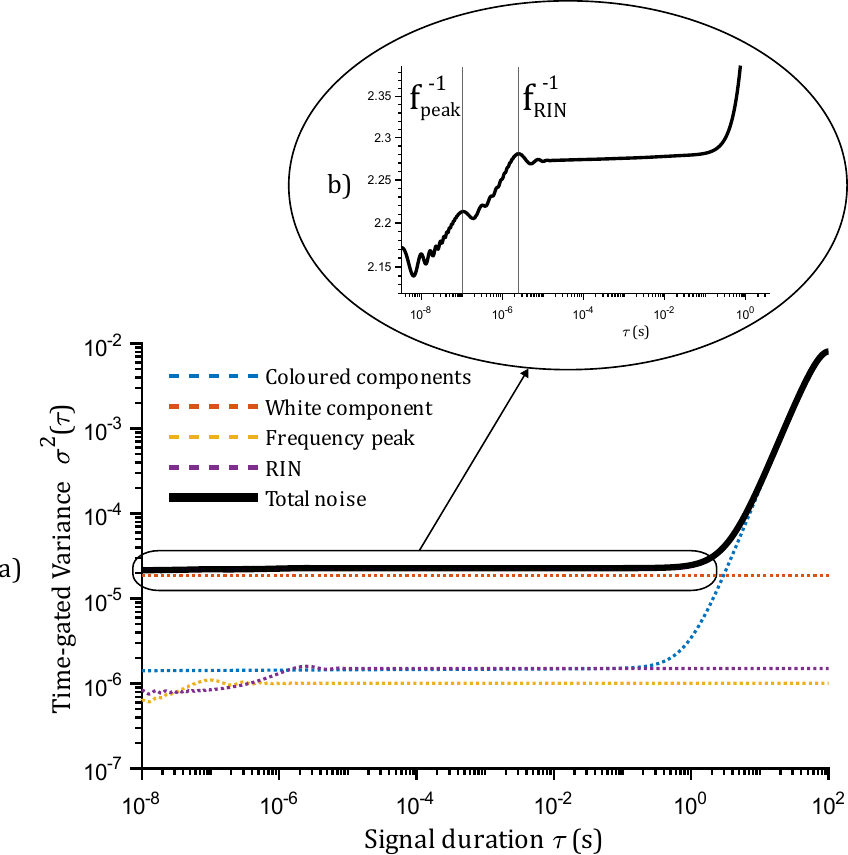}
    \caption{\acrshort{TGV} as a function of calibration duration. (a) \acrshort{TGV} of the full receiver noise and its individual contributions. (b) Zoom on the total receiver noise, highlighting the step-like effect of \gls{RIN} and the frequency spike on the \acrshort{TGV}.}

    \label{fig:ft_var_zoom}
\end{figure}

Each noise contribution in this model results in a distinct \acrshort{TGV} that depends on the calibration duration, as shown in Figure \ref{fig:ft_var_zoom}. The zoomed-in view highlights the impact of modeled imperfections such as \gls{RIN} and frequency spikes, which cause a sudden increase in the \acrshort{TGV} estimate when the calibration duration exceeds the inverse of the corresponding impaired frequency.  This behavior is quite intuitive, as shorter calibration times effectively filter out slower noise sources, preventing them from impacting the variance estimate. A similar analysis is conducted for the electronic noise (not shown), which is combined with the total receiver noise to estimate the shot noise variance.

\glsreset{SNC}
\subsection{Impact of calibration duration on characterized dynamics \gls{SNC}}
\label{section:calib_duration}
As discussed in section \ref{section:conventional}, the typical approach for estimating shot noise variance is to subtract the electronic noise variance from the full receiver noise variance estimate. We can now express the calibration-time-dependent shot noise variance estimator as:

\begin{equation} \hat{N_0}(\tau)= \sigma_{\text{rec}}^2(\tau) -\sigma_{\text{elec}}^2(\tau) = N_0 + \delta_{\text{rec}}(\tau) \label{eq:model_shot} \end{equation}

\begin{figure}
    \includegraphics[width=0.48\textwidth, trim= 0 0 0 0, clip]{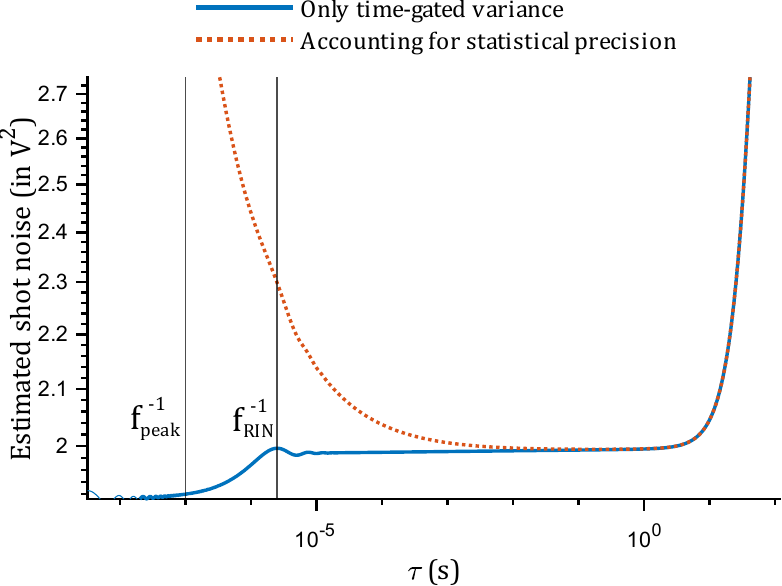}
    \caption{Estimated shot noise variance as a function of calibration time. The dashed line attempts to account for statistical precision using worst-case estimators, while the solid line does not.}

    \label{fig:shot_estim_v2}
\end{figure}

This estimator is depicted as the solid curve in figure \ref{fig:shot_estim_v2}, where we observe that the shot noise variance estimate increases with calibration time. Two effects contribute to this increase. One is due to the \gls{RIN} (as seen in figure \ref{fig:ft_var_zoom}), and the other, which becomes more significant at longer calibration durations, is due to additional low-frequency components introduced by the \gls{LO} laser. Interestingly, the frequency spike observed in the spectrum does not affect the shot noise estimate at all. This is because the spike is solely associated with the detection electronics, and is thus present in both receiver and electronic noise contributions.

To account for statistical precision in variance estimation, one approach is to compute worst-case estimators \cite{roumestan_advanced_2022} (see appendix \ref{app:worst_case_estimator}). This involves estimating the maximum and minimum possible values of the variance estimator and selecting the least favorable for security purposes. In this figure, the dashed curve illustrates the potential increase in variance resulting from a reduced number of samples due to shorter calibration durations, while maintaining a fixed sampling rate of $625$ MHz (matching our experimental setup). This underscores the trade-off between calibration time and the reliability of the variance estimate.

The clear minimum observed in figure \ref{fig:shot_estim_v2}, in conjunction with our model from equation \eqref{eq:model_shot}, provides insight into the question: "How long should we calibrate to obtain the most accurate shot noise estimate using characterized dynamics \gls{SNC}?". To achieve the most precise estimate, it is essential to minimize the positive $\delta_{rec}(\tau)$ component, thereby bringing the estimator as close as possible to the true shot noise variance $N_0$. With the chosen parameters, this optimum occurs at approximately $10^{-1}$ seconds.

\glsreset{SNC}
\section{Characterized dynamics fully white \gls{SNC}}
\label{sec:cdfwsnc}
In this section, we propose a novel method that leverages noise spectral properties to achieve a more precise and robust calibration of shot noise. This approach aims to produce a variance estimate that is closer to the true shot noise $N_0$ and less sensitive to calibration duration. In particular, it exploits the known spectral properties of true shot noise that is white. However, this improvement comes at the cost of introducing additional assumptions to compute the estimate.

\subsection{Extra assumption} This model assumes prior knowledge in the form of the following assumption : the only sources of white noise present in the receiver noise are shot noise and electronic noise. Therefore, the full receiver white noise variance can be expressed as: \begin{equation} \sigma_{\text{rec,white}}^2 = \sigma_{\text{elec,white}}^2 + N_0 \end{equation}

\subsection{Noise Model and Shot Noise Estimator}
Using these assumptions, we get the following equations:

\begin{equation} \sigma_{\text{elec}}^2(\tau) = \sigma_{\text{elec,white}}^2 + \sigma_{\text{elec,not white}}^2(\tau) \end{equation}
\begin{equation} \sigma_{\text{rec}}^2(\tau) = \sigma_{\text{elec,white}}^2 + \sigma_{\text{elec,not white}}^2(\tau) + N_0 + \delta_{\text{rec}}(\tau) \end{equation}

From which we can derive the shot noise estimator:

\begin{equation} \hat{N_0} = \sigma_{\text{rec,white}}^2 - \sigma_{\text{elec,white}}^2 \end{equation}

\glsreset{SNC}
\subsection{Implications and Comparison with characterized dynamics \gls{SNC}}
The characterized dynamics fully white \gls{SNC} method, allows us to define the shot noise level under the assumption that the only sources of white noise in the receiver are electronic and shot noise. This provides a reference shot noise value against which we can compare the results obtained with the characterized dynamics \gls{SNC} method.

In particular, by relying on the characterized dynamics fully white \gls{SNC}, we can determine the "true" \gls{SNU} and express the relative calibration drift $\delta_{rec}(\tau)$, as estimated by the characterized dynamics \gls{SNC}, in units of this true \gls{SNU} over the calibration time. This enables us to quantify the misestimation introduced by imperfect shot noise calibration in realistic experimental conditions.

\begin{figure}
    \includegraphics[width=0.48\textwidth, trim= 0 0 0 0, clip]{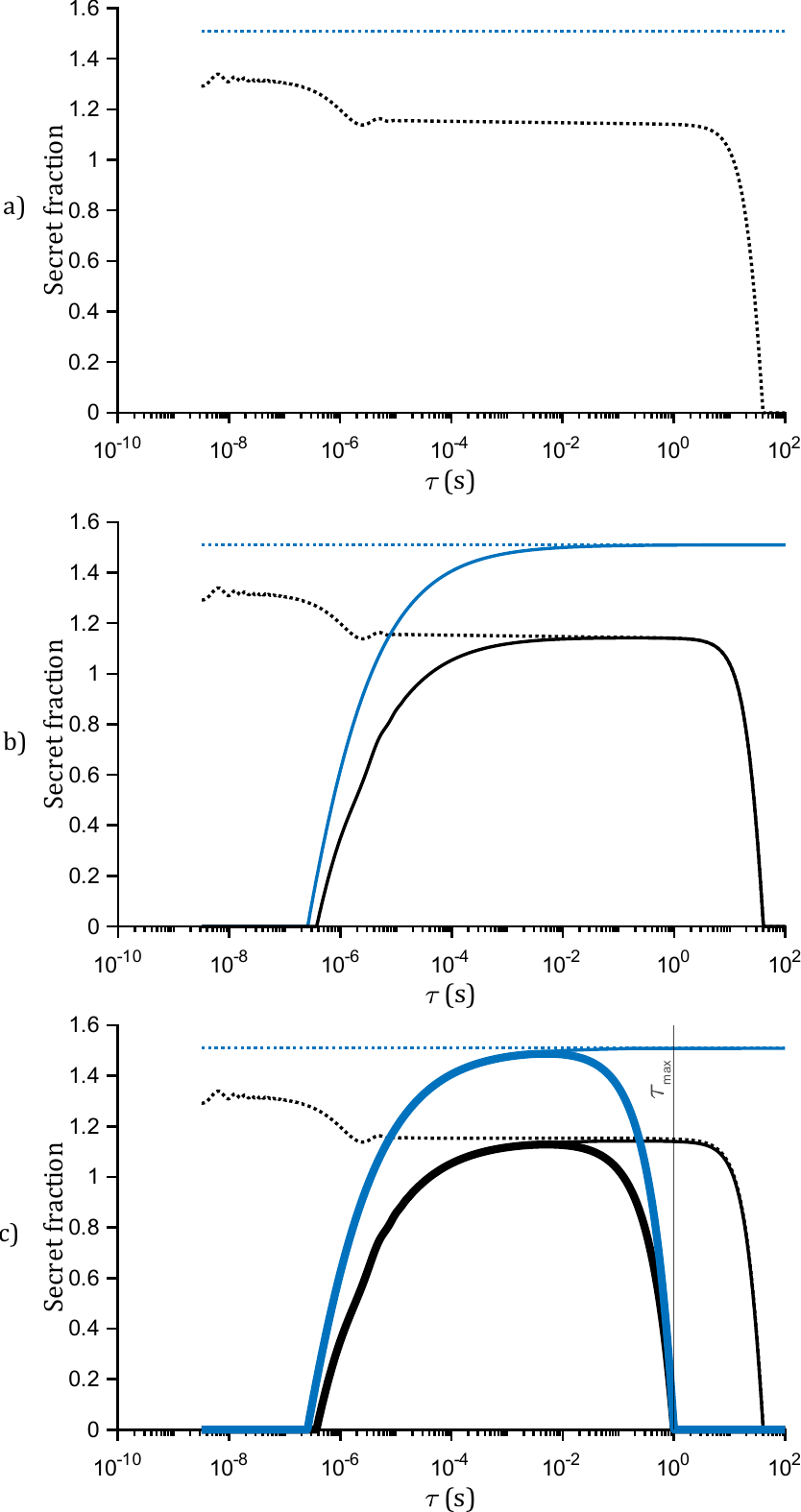}
\caption{Secret key fraction as a function of calibration time. This figure illustrates the trade-offs involved in choosing the calibration duration: (a) time-gated calibration of shot noise increase with calibration duration; (b) shorter calibration durations reduce statistical precision at a fixed sampling rate; and (c) calibration consumes time that could otherwise be used for key generation, reducing the effective key rate via a duty-cycle factor. This factor depends on the maximum calibration duration allowed, for instance, the \gls{WSS} timespan called here $\tau_\text{max}$. Black curves correspond to characterized dynamics \gls{SNC}, and blue curves to characterized dynamics fully white \gls{SNC}.}
\label{fig:SKR}
\end{figure}

To assess the impact of this misestimation in a practical case, we consider an otherwise ideal \gls{CV-QKD} experiment. The parameters used for this study are listed in Table \ref{tab:params}. 
\begin{table}

\centering
\caption{Ideal parameters used to assess the impact of calibration errors.}
\begin{tabular}{c|c}

\textbf{Parameter} & \textbf{Value} \\
\hline
\hline
Electronic noise (\gls{SNU})  & 0.09 \\
Modulation variance $V_A$ (\gls{SNU})  & 5 \\
Excess noise measured at Bob's side (\gls{SNU})  & 0.025 \\
Transmittance T  & 0.25 \\

\end{tabular}
 % Adjust space here as needed
\label{tab:params}
\end{table}
These parameters allow us to compute the \gls{SKR} of this nearly ideal \gls{CV-QKD} scenario using both our proposed characterized dynamics fully white \gls{SNC} (under the assumption that the model hypotheses hold) and a characterized dynamics \gls{SNC}. An overview of the respective noise models and estimators is given in Table~\ref{tab:overview}, and the corresponding secret key fractions are compared in Figure~\ref{fig:SKR}. We rely on the analytical expression of the asymptotic secret key rate for Gaussian-modulated \gls{CV-QKD}.

This comparison demonstrates that the secret fraction can be improved with the characterized dynamics fully white \gls{SNC}. Interestingly, it also shows greater tolerance to detection imperfections. For instance, the impact of \gls{RIN} is observable in characterized dynamics \gls{SNC}, whereas it has no impact in  characterized dynamics fully white \gls{SNC}.

Figure \ref{fig:SKR}-c clearly illustrates the impact of the $\tau_\text{max}$ constraint. In the absence of this limitation, the results would correspond to those shown in figure \ref{fig:SKR}-b. To determine the optimal calibration duration, we must analyze as is done in figure \ref{fig:SKR} the trade-off between:
\begin{itemize}
    \item \textbf{Statistical precision}: More samples improve variance estimation, which prevents excessively short calibrations.
    \item \textbf{Impact of calibration duration on variance estimation}: Depending on noise spectral properties, calibration time can significantly affect variance estimation. Some noise contributions may become dominant at certain time scales while remaining negligible at others.
    \item \textbf{Dead time}: During calibration, no key exchange occurs. This introduces an efficiency correction in a full experiment, represented by a simple multiplicative factor on the key rate. If the signal is used with a duty cycle for key exchange, the secret key rate must be multiplied by this factor to account for the system's dead time. This prevents excessively long calibrations.
\end{itemize}

\begin{figure}
    \includegraphics[width=0.48\textwidth, trim= 0 0 0 0, clip]{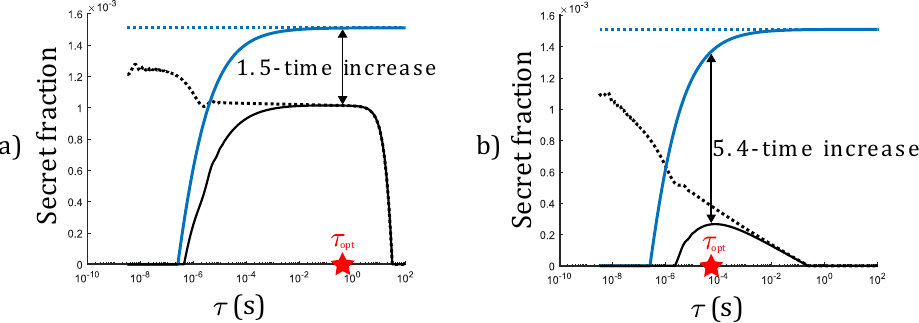}
    \caption{Secret key fraction as a function of calibration time. (a) Low 1/f noise. (b) higher 1/f noise (40-fold increase). Black curves correspond to characterized dynamics \gls{SNC}, red stars indicate the optimal calibration time $\tau_\text{opt}$, and blue curves correspond to characterized dynamics fully white \gls{SNC}.}
    \label{fig:SKR1}
\end{figure}
 
Figure \ref{fig:SKR1} illustrates the secret key fraction as a function of calibration time in two scenarios: a) A low \gls{RIN} LO laser and b) a higher \gls{RIN} LO laser, where the increased \gls{RIN} leads to a 40-fold increase in the 1/f noise component. These results indicate that practical gains depend on the quality of the receiver and \gls{LO}, with realistic scenarios showing significant performance improvements, as seen in figure \ref{fig:SKR1}-b, where the \gls{LO} induced excess noise contributes notably at longer calibration durations.

\section{Experimental validation}
\subsection{Experimental Setup}
\label{sec:exp_model}
Our experimental setup is an adapted version of that described in \cite{aymeric_quantum_2022, aymeric_symbiotic_2022}. The \gls{LO} is provided by an NKT Koheras Adjustik laser with a 100 Hz linewidth. Optical mixing is performed using a Kylia COH28-X dual-polarization 90° hybrid, followed by \gls{BHD} for each quadrature measurement using Exalos EBR370005-02 detectors with a 250 MHz bandwidth. The detected signals are then digitized and stored using a Tektronix MSO44 oscilloscope, operating at a sampling rate of 625 MHz. While this setup is part of a complete \gls{CV-QKD} system, only the detection stage is relevant for the present study.

\glsreset{WSS}
\subsection{\gls{WSS}}
\label{test WSS}
To rigorously assess \gls{WSS} in \gls{CV-QKD} noise calibration, we employ the methodology proposed by Zhivomirov and Nedelchev \cite{zhivomirov_method_2020}. This method tests whether statistical properties remain consistent over time by splitting a signal into two segments and applying:
\begin{itemize}
    \item \textbf{Wilcoxon rank-sum test}: Checks for mean equality.
    \item \textbf{Brown–Forsythe test}: Evaluates variance stability.
    \item \textbf{Autocovariance function comparison}: Measures similarity in temporal correlation structures.
\end{itemize}

A signal passing all three tests is classified as \gls{WSS}. If deviations occur, the assumption is invalid, necessitating adjustments in calibration procedures.

To evaluate the stationarity properties of the experimental noise signals, we analyze a dataset consisting of 268 independent blocks. Each block contains \(10^6\) samples, collected using the experimental setup described in Section~\ref{sec:exp_model} with the configuration ``LO on, signal off.'' This corresponds to a time duration of \(\tau_{\text{block}} = 1.6\,\mathrm{ms}\) per block.

\begin{figure}
    \centering
    \includegraphics[width=0.48\textwidth]{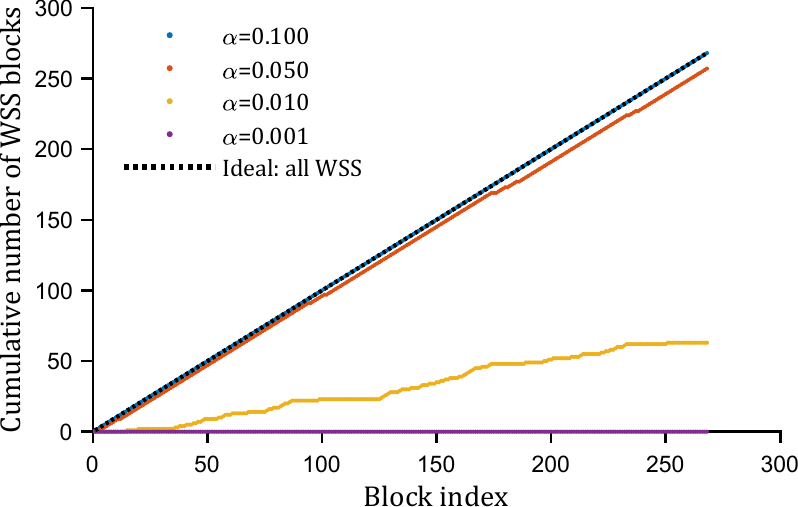}
    \caption{Cumulative number of blocks identified as \gls{WSS} versus block index for different confidence levels \(\alpha\). The dataset comprises 268 blocks of \(10^6\) samples each (\(\tau_{\text{block}}=1.6\,\mathrm{ms}\)) recorded with "\gls{LO} on, signal off".}
    \label{fig:WSStest}
\end{figure}

Figure~\ref{fig:WSStest} presents the cumulative number of blocks classified as Wide-Sense Stationary (\gls{WSS}) plotted against the block index, for several confidence levels \(\alpha\). For relatively lenient confidence thresholds (\(\alpha > 0.05\)), nearly all blocks satisfy the \gls{WSS} criterion (100\% for \(\alpha=0.1\) and 95.9\% for \(\alpha=0.05\)). Stricter significance levels result in fewer blocks passing the test: none for \(\alpha=0.001\) and only 23.5\% for \(\alpha=0.01\).

These results demonstrate, with reasonable confidence, that the experimental noise signals can be considered \gls{WSS} over the timescales relevant to this study, validating the applicability of our stationarity test for the durations typical of our measurements.

%\glsreset{TGV}
\subsection{\acrshort{TGV}}

When inserting the parameters obtained from experimental data fits, the \acrshort{TGV} of receiver noise defined in Equation \eqref{sigma gated} is expected to increase as the calibration duration increases (see appendix \ref{ann:ex}).

However, verifying this trend experimentally presents challenges. It is not feasible to obtain comparable datasets for different durations while maintaining a fixed sampling rate. Keeping the sampling rate constant would require varying the number of data points, affecting statistical precision and potentially hitting memory constraints for long calibration durations. Conversely, adjusting the sampling rate can alter the oscilloscope’s gain levels (especially the electronic bandwidth of the oscilloscope that cannot be tuned at will), complicating direct comparisons. To overcome these limitations, we employed a numerical downsampling approach to investigate whether the expected variance increase appears over longer durations.

Our strategy was to start with a long measurement (20M samples, corresponding to a 40s acquisition at 0.5 MHz) and then generate smaller datasets, each representing a different duration but containing the same number of samples, $N$. This was achieved by digitally downsampling the signal to match the target duration while keeping $N$ fixed. A critical challenge in this process is aliasing, which causes high-frequency noise components to leak into the downsampled signal, distorting the variance estimate. To counteract this, we applied a low-pass filter matched to the new effective sampling rate before downsampling. While this filtering reduces the variance estimate (effectively lowering the upper integration limit in equation \eqref{sigma gated}), it successfully eliminates aliasing, enabling a clearer observation of the \acrshort{TGV} effect. However, this additional filtering complicates direct comparisons with theoretical expectations that do not account for such filtering effects.

\begin{figure}
    \includegraphics[width=0.48\textwidth, trim= 0 0 0 0, clip]{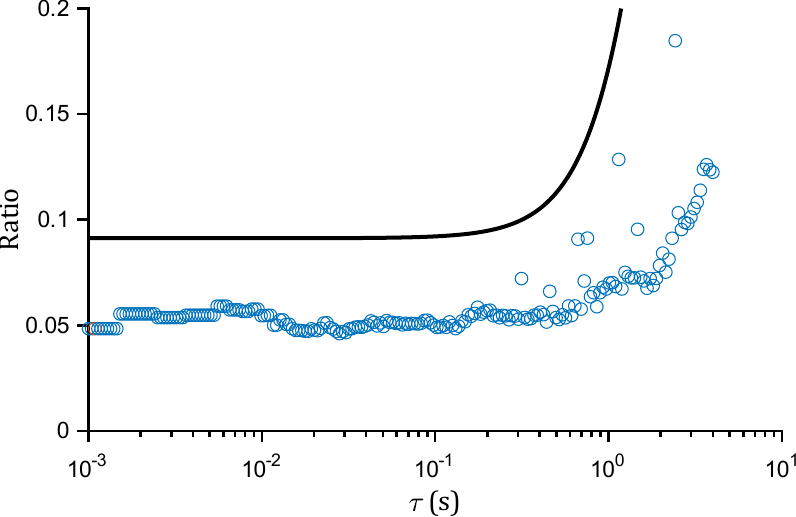}
    \caption{Ratio of electronic noise to full receiver noise as a function of calibration duration. The black curve represents the theoretical \acrshort{TGV} derived from experimentally fitted \gls{PSD} models, while the circles correspond to variance estimates obtained from experimental signals of different durations.}

    \label{fig:var_elec_loc}
\end{figure}

To mitigate this issue, we instead examine the ratio of electronic noise to full receiver noise, as illustrated in Figure \ref{fig:var_elec_loc}. This ratio provides a qualitative validation of the expected variance increase, demonstrating the agreement between theoretical predictions and experimental results.

\section{Conclusion}

In this paper, we have addressed fundamental limitations in traditional noise calibration for \gls{CV-QKD} systems by explicitly accounting for and characterizing noise variance dynamics, which are temporal fluctuations of the noise variance over finite observation windows. We have developed a comprehensive framework that encompasses a critical examination of the stationarity assumption and its role in ensuring calibration validity, supported by \gls{WSS} testing to justify this assumption, a method to determine the optimal calibration duration based on the concept of \acrshort{TGV}, and a novel technique to isolate the shot noise component using white noise isolation techniques. Our results demonstrate that this approach significantly enhances performance and robustness against receiver imperfections, including those related to \gls{LO} fluctuations and detection electronics. These advancements pave the way toward more practical and cost-effective \gls{CV-QKD} implementations.

Importantly, a central contribution of this work is also the formalization of the calibration procedures themselves, which are integral to \gls{CV-QKD} protocols. By explicitly clarifying previously implicit assumptions and establishing a rigorous calibration framework, we lay essential groundwork for the certification of these procedures. Such certification is crucial for \gls{CV-QKD}, as it constitutes a key step toward the reliable deployment of this technology in real-world, secure quantum communication networks.

Future work will consider integrating digital signal processing techniques within the \acrshort{TGV} framework, further refining calibration strategies and system performance.

\section*{Acknowledgements}

This project has received funding from the European Union’s Horizon Europe research and innovation program under the project Quantum Security Networks Partnership (QSNP, grant agreement No. 101114043). The authors also acknowledge support from the French Quantum PEPR program, project QCommTestbed. Guillaume Ricard moreover gratefully acknowledges the support of the Paris Region PhD program.

\appendices

\section{Statistical precision and worst case estimators}
\label{app:worst_case_estimator}

One way to account for statistical precision in variance estimation \cite{roumestan_advanced_2022} is to choose an $\epsilon_{\text{secu}}$ parameter such that the estimators are $\hat{V}_{max}$ or $\hat{V}_{min}$, depending on if over or under estimation is most advantageous to Eve. We have:

\begin{align}
    P( \hat{V}_{max} \geq V) &\geq 1-\epsilon_{\text{secu}} \\
    P( \hat{V}_{min} \leq V) &\geq 1-\epsilon_{\text{secu}}
\end{align}

And so:

\begin{align}
    \hat{V}_{max} &= \hat{V} + \sigma Q_{1-\epsilon_{\text{secu}}} \\
    \hat{V}_{min} &= \hat{V} - \sigma Q_{1-\epsilon_{\text{secu}}}
\end{align}

With $Q_{1-\epsilon_{\text{secu}}}$ the ${1-\epsilon_{\text{secu}}}$ quantile of a centered and normalized Gaussian distribution.

We can then assimilate $\sigma$ to the statistical uncertainty over the estimation:

\begin{align}
    \sigma &= \frac{\hat{V}}{\sqrt{N}}
\end{align}

We finally have:

\begin{align}
    \hat{V}_{max} &= \hat{V} \left(1 + \frac{1}{\sqrt{N}} Q_{1-\epsilon_{\text{secu}}} \right) \\
    \hat{V}_{min} &= \hat{V} \left(1 - \frac{1}{\sqrt{N}} Q_{1-\epsilon_{\text{secu}}} \right)
\end{align}


\section{Example: \acrshort{TGV} Estimation with a Simple \gls{PSD} Model}  
\label{ann:ex}

\subsection{Noise Model \gls{PSD}}  

To illustrate the analytical computation of finite-time variance, we consider a simple example using a generic power-law noise model for the \gls{PSD}:  

\begin{equation}
\label{PSD_model}
S_x(f) = \sum_{\alpha = -2}^{2} h_{\alpha} f^{\alpha}, 
\end{equation}

where each term corresponds to a power-law noise component with exponent $\alpha$ and amplitude $h_{\alpha}$. This type of model is widely used in frequency metrology, as it captures common noise processes such as white noise, flicker noise, and random walk noise \cite{makdissi_stability_2009}.  

While this framework is general, the goal here is to present a numerical example to illustrate how the finite-time variance can be computed analytically. Additional noise components, such as \gls{RIN} and frequency spikes, are introduced in Section \ref{results} to further refine the model.  

\subsection{Computation of the Autocovariance Function}  

The inverse Fourier transform of equation \eqref{PSD_model} gives the corresponding autocovariance function $R_x(t)$. Using the linearity of the Fourier transform, we compute each term separately:  

\begin{align}
    \mathcal{F}^{-1}(S_x(f))(t) &= \sum_{\alpha = -2}^{2} h_{\alpha}  \int_{-\infty}^{\infty} f^{\alpha} e^{2i\pi f t}  df \\
    &= R_x(t).
\end{align}

For this example, we obtain:  

\begin{align*}
    R_x(t) &=  -2 \pi^2  \,{h}_{-2} \,t\,\mathrm{sign}\left(t\right) + \mathrm{i} \pi \,{h}_{-1} \,\mathrm{sign}\left(t\right)\\
    &\quad+h_0 \,{\delta }\left(t\right)-\frac{ \mathrm{i} h_1 \,{{\delta }}^{\prime } \left(t\right)}{2\,\pi } -\frac{h_2 \,{{\delta }}^{\prime \prime } \left(t\right)}{4\,\pi^2 }.
\end{align*}

This explicit form allows us to examine how different noise components contribute to the autocovariance.

\subsection{Effect of Finite-Time Gating on the Power Spectral Density}  

To analyze the effect of finite observation time $\tau$, we compute the gated \gls{PSD}, $S_{x,\text{gated}}(f,\tau)$:

\begin{align}
    S_{x,\text{gated}}(f,\tau) &=  \sum_{\alpha = -2}^{2} I_{\alpha}(f,\tau)
\end{align}

with $I_{\alpha}(f,\tau)$ corresponding to the Fourier transform of the components of the time-gated autocorrelation function. It is easy to check that time-gating has no effect on components $\alpha = 0,1,2$. We get :

\begin{align}
    S_{x,\text{gated}}(f,\tau) &=  I_{-2}(f,\tau) + I_{-1}(f,\tau) + h_0 + f h_1 + f^2 h_2
\end{align}

where the terms $I_{-2}(f,\tau)$ and $I_{-1}(f,\tau)$ account for the time-limited observation effects on low-frequency components:

\begin{align}
    I_{-2}(f,\tau) &= \int_{-\tau/2}^{\tau/2} -2 \pi^2  \,{h}_{-2} \,t\,\mathrm{sign}\left(t\right) e^{-2i\pi f t} dt \\
    &= -\frac{h_{-2}(\pi f \tau \sin(\pi f \tau) + \cos(\pi f \tau) -1)}{f^2},
\end{align}
\begin{align}
    I_{-1}(f,\tau) &= \int_{-\tau/2}^{\tau/2}   \,\mathrm{i} \pi {h}_{-1} \,\mathrm{sign}\left(t\right) e^{-2i\pi f t} dt \\
    &= - \frac{h_{-1}(\cos(\pi f \tau) -1) }{f}.
\end{align}

\section{Identification of the White Noise Component}

The effectiveness of the Characterized-dynamics fully white \gls{SNC} method hinges on accurately isolating and estimating the white noise components within the signals. Automating this process remains a challenging task, particularly in situations where the noise structure is only partially characterized. In this context, two complementary methodological approaches have been identified and preliminarily explored. Both are presented here as potential strategies to enhance the robustness and generality of the method.

\subsection{ARMA Models}

A first approach relies on the use of \gls{AR} \gls{MA} models, widely used in time series analysis
for modeling and forecasting signals. They combine two
components:
\begin{itemize}
    \item An \textbf{\gls{AR} (Autoregressive)} component, which expresses the current value as a linear combination of its past values.
    \item An \textbf{\gls{MA} (Moving Average)} component, which models the current value as a linear combination of past white noise terms.
\end{itemize}

In essence, an ARMA model can be seen as a feed-forward
and feedback filter \cite{kobayashi_probability_2011} that takes white noise as input and
outputs a signal resembling the original data. By fitting an
ARMA model to our signal, we can approximate the relationship between the observed data and the underlying white noise. Once the model is fitted, the residuals (the difference between
the observed data and the model’s prediction) can provide a
close estimate of the isolated white noise component. This method presents the advantage of being fully data-driven and does not require prior knowledge of the noise variance or spectral content.

\subsection{Spectral Estimation and Wiener Filtering}

A second approach exploits frequency-domain techniques to identify the white noise component from the total signal. When the \gls{PSD} of a signal is estimated, the white noise contribution typically appears as a flat baseline. In practice, however, the presence of colored noise components can render the direct identification of this baseline nontrivial.

A pragmatic strategy consists in estimating the white noise level by selecting the minimum value of a smoothed PSD estimate, under the assumption that at least one frequency band remains predominantly white. Based on this estimate, a Wiener filter can be constructed to attenuate the colored noise contributions relative to the white noise level:
\begin{equation}
    H(f) = \frac{P_{\text{white}}}{P_{\text{signal}}(f)} = \frac{P_{\text{white}}}{P_{\text{white}} + P_{\text{coloured}}(f)}
\end{equation}
where \( P_{\text{white}} \) is the estimated white noise power and \( P_{\text{signal}}(f) \) is the measured power spectral density.

This filter can then be applied in the frequency domain to extract an estimate of the isolated white noise contribution, either through direct inverse Fourier transform of the filtered spectrum or by frequency-domain filtering of the original signal.

\bibliographystyle{IEEEtran}
\bibliography{Ma_bibliotheque_clean}

\end{document}